# Free-space quasi-phase matched second harmonic generation in crystalline quartz

NAZAR KOVALENKO[*,1], ANKIT PAI[1] AND OLEG PRONIN[1]

[1]Helmut Schmidt University, Holstenhofweg 85, 22043 Hamburg, Germany
*Corresponding author: nazar.kovalenko@hsu-hh.de

**Abstract:** We report experimental results on second-harmonic generation in a z-cut quartz crystal under conditions of *free-space quasi-phase matching* in a multi-pass cell. In a 62-pass configuration, an efficiency of 0.027% or $1.4 \cdot 10^{-4}$ %/MW/cm$^2$ was achieved, delivering 1 uJ of the second harmonic at 3.7 mJ pump pulse. This corresponds to an enhancement factor of more than 1000 in conversion efficiency as compared to a single pass. The generated second-harmonic beam demonstrates high beam quality M$^2$=1.1 and linear polarization. The scaling of the output power with the number of passes is in good agreement with the calculated values. Further increasing the pump intensity, number of passes, and amount of plates opens the way to scaling the conversion efficiency to values on the order of tens of percent.

## Introduction

Since the first demonstration of nonlinear optical frequency conversion by Peter Franken [1], crystalline quartz has remained a nonlinear optical material that continues to attract the attention of researchers today. Quartz lacks true (birefringent) phase matching and has a relatively low nonlinear susceptibility of 0.3 pm/V; however, its transparency in the vacuum ultraviolet (VUV) up to 150 nm and its high damage threshold (900 GW/cm$^2$ [2] motivate researchers to explore various approaches to increasing the efficiency of nonlinear optical conversion based on the principles described in [3]. In the works [4,5], the authors created a periodic structure analogous to those in periodically poled ferroelectric crystals by applying mechanical stress to a quartz crystal. In the work [6], a periodic structure was created by alternating thin plates of crystalline quartz. The initially identical plates were joined with a periodic change in crystallographic orientation. As a result, a nonlinear dependence of the efficiency of nonlinear optical conversion on the number of layers was observed, which is typical for quasi-phase matching. In the work [7], the authors created a periodic structure in crystalline quartz by modifying the optical properties of the medium using ultrashort laser pulses. The effect is achieved by "switching off" the second-order nonlinear susceptibility of the nonlinear medium through disordering of the crystalline structure in a localized region subjected to laser irradiation. In the above-mentioned approaches, the fundamental beam and the second harmonic propagate through the entire structure, experiencing the influence of dispersive properties of the medium, such as group-velocity mismatch and nonlinear dispersion. Previously, we demonstrated [8] an alternative method for quasi-phase-matching based on multipass cells and experimentally demonstrated its proof-of-principle operation. In the proposed method, each pass through the cell is equivalent to a layer of a periodic structure. Moreover, the scheme fundamentally allows the phase of the pump beam and the second harmonic to be adjusted after each pass. This method was recently extended into parametric amplification of the femtosecond pulses [9]. Here, as a next step, to prove the scalability of the proposed FSQPM approach, we present experimental results on second-harmonic generation in a quartz crystal using a Herriott-type cell with an increased number of passes and anti-reflection-coated crystalline quartz.

## Setup and experimental results

The pump source is a nanosecond laser with up to 10 mJ andpulse duration of 6 ns (Quantum Light Instruments, Q1B-10-1064), the same unit was used in [8]. The optical setup was modified to increase the number of passes from 14 in our previous work to 62 (fig. 1). The cell was formed by a pair of concave mirrors with a diameter of 25 mm and radii of curvature of 500 mm. The relatively short cell, 194 mm, allowed a large number of beam passes to be accommodated. A z-cut AR-coated crystalline quartz plate with a diameter of 25 mm and a thickness of 3 mm was used as the nonlinear medium. The anti-reflection coatings had maximum transmission at the pump and second-harmonic wavelengths. The number of passes was limited by the beam size on the cell's mirrors and by the geometric dimensions of the mirrors. The transmission of the cell with the full set of optical elements, including the curved mirrors, a quartz plate (CQ, Fig. 1a), and an AR-coated fused silica phase-compensating plate (FS, Fig. 1a), was 82% at the fundamental wavelength and 59% at the second-harmonic wavelength. The transmission is mainly determined by reflection losses in the quartz and the phase compensator.

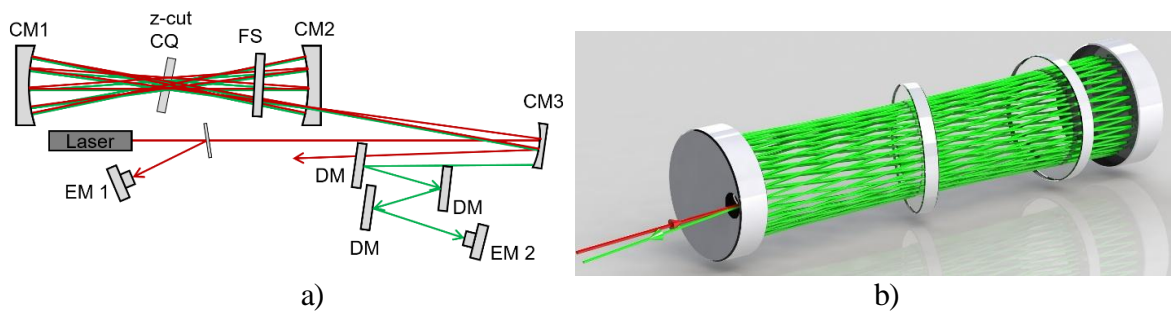

*Fig.1 Experimental setup: CM1, CM2-curved mirrors, z-cut CQ –crystalline quartz plate, FS-fused silica plate for the phase correction, CM3 – curved mirror for the mode matching, DM – dichroic mirrors, EM1, EM2 – pyroelectric energy meters.*

We used only a single-phase compensator to reduce reflection losses. The available compensator adjusted the phase in one arm of the optical setup. The relative phase in the other arm was adjusted by translating the crystalline plate along the cell's optical axis. In this case, phase adjustment occurred due to a change in the optical path length in the gaseous medium (air) inside the cell and, consequently, a change in the phase difference between the fundamental beam and the second harmonic due to the natural dispersion of the gas medium. Further reduction of losses is possible by omitting solid-state phase compensators. In this case, phase tuning in both arms can be achieved by varying the pressure of the gaseous medium in the cell within a small range, on the order of 1 bar. The drawback of this approach is the need to place the optical setup in a sealed enclosure capable of varying and maintaining a stable pressure.

The output pulse energies of the second harmonic and the fundamental radiation reported below are given "as measured," without correction for losses in the cell. The input beam was single-mode, had an $M^2$ value of 1.2, and exhibited a noticeable level of astigmatism (Fig.2a). The output pump beam retained its astigmatism and showed an $M^2$ value of 1.2.

The output second-harmonic beam showed higher beam quality, with $M^2 = 1.1$, and significantly lower astigmatism (Fig. 2b).

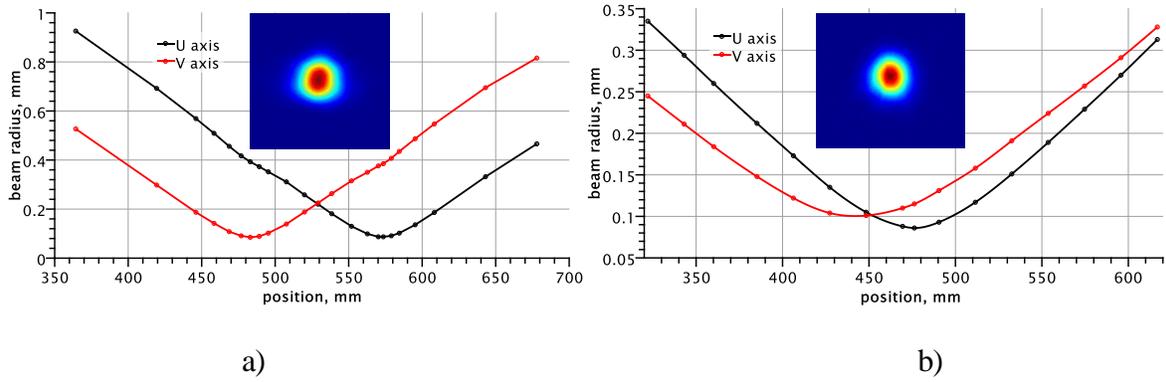

*Fig.2 The input pump beam (a) and second harmonic beam (b) characterization (conforms ISO 11146-1): Gaussian beam caustic with the beam profile at the waist presented in the inset.*

The thickness of the crystalline quartz plate should ideally be equal to an odd number of coherence lengths. Therefore, to increase the conversion efficiency, we experimentally adjusted the angle of incidence. At an angle of 19° between the optical axis of the cell and the normal to the plate, the maximum output energy of the second harmonic was obtained. Such an angle of incidence corresponds to 149 coherence lengths along the optical path in crystalline quartz. A second objective of adjusting the angle of incidence was to reduce the gyration effect in the crystalline quartz [5]. At the defined optimal angle of incidence, the gyration effect for the fundamental beam was negligible.

The angle of incidence was adjusted in the plane of polarization of the pump beam, which allows consider the pumping as extraordinary-wave excitation. The plane of polarization of the second harmonic was orthogonal to that of the pump, indicating the predominance of an ee-o type nonlinear interaction.

The efficiency of the nonlinear conversion is defined as the ratio of the pulse energy of the second harmonic to the pump pulse energy at the entrance to the multipass cell. The Q-switched laser used did not employ longitudinal mode selection; as a result, the temporal profiles of the pump pulses and the second-harmonic pulses consisted of a sequence of sub-nanosecond spikes (Fig. 3). It turns out that this pulse structure is typical of Q-switched nanosecond lasers. The measurements were performed using a Keysight DSOS254A oscilloscope with 2.5 GHz bandwidth and photodiodes with a pulse rise time of less than 175 ps. Despite the spiky structure of the pulses, approximating the pulse with a Gaussian curve yields a sufficiently stable full width at half maximum of $9.3 \pm 0.15$ ns. The pulse structure brings uncertainty in measuring the conversion efficiency. Therefore, to estimate the energy of the second-harmonic pulses, we accumulated measurements of the second-harmonic pulse energy at a constant pump energy. The resulting distribution is shown in the inset of Fig. 3. To suppress the residual pump beam, triple filtering with dichroic mirrors was implemented, providing an attenuation factor of approximately $8.8 \times 10^{-8}$, corresponding to a residual fundamental energy of about 0.3 nJ. At a pump energy of 3.7 mJ, the reached output pulse energy of the second harmonic was 1 µJ. With a measured second-harmonic energy of approximately 0.7 nJ per single pass conversion, the enhancement factor in the multipass scheme is on the order of 1000. The fundamental beam diameter at the crystalline quartz plate was 530x510 µm, which, for a FWHM of the pulse duration of 9.3 ns, corresponds to a pump-beam intensity of 190 MW/cm². Thus, the conversion efficiency of the pump radiation into the second harmonic is 0.027% or 0.00014% per MW/cm².

The theoretical conversion efficiency in the multipass cell was estimated using a formula valid in the SI system of units, describing the energy transfer between the pump beam and the second harmonic without pump depletion.

$$\eta(z) = \frac{2^3 \cdot \pi^2 \cdot d^2}{n_\omega^2 \cdot n_{2\omega} \cdot c \cdot \lambda^2 \cdot \varepsilon_0} I \cdot z^2 \cdot sinc^2\left(\frac{\Delta k \cdot z}{2}\right),$$

where d-nonlinear coefficient, $n_\omega$, $n_{2\omega}$ - refractive index, $\lambda$-fundamental wavelength, $\varepsilon_0$ – vacuum permittivity, I – pump intensity, $\Delta k$ – wave-vector mismatch, z - coordinate.

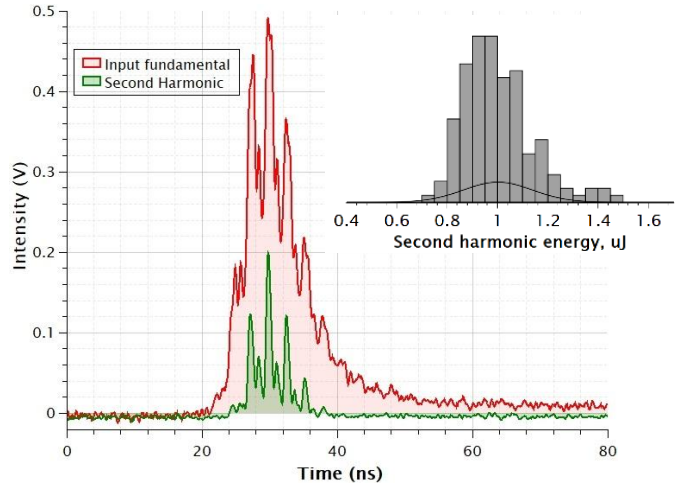

*Fig.3 Temporal profile of the fundamental Q-switched pulses (red) at 1064 nm and generated second harmonic (green). Inset – second harmonic energy distribution for the same pump energy accumulated over 150 shots.*

Assuming a coherence length for eeo interaction type at given angle of incidence $l_c = \frac{\pi}{\Delta k}$ of 21.5 μm and a pump intensity of 190 MW/cm², we obtain an estimated conversion efficiency within a single coherence length of $2.3 \times 10^{-7}$. Taking into account the number of passes N=62, the calculated conversion efficiency is $\eta(l_c) \cdot N^2 = 0.087\%$, or 0.00046%/MW/cm². The theoretical estimate exceeds the experimental efficiency by a factor of ~3.3. We attribute this deviation to overall losses in the anti-reflection coatings and to inhomogeneous phase-matching conditions arising from variations in the angles of incidence of the interacting beams during different passes through the cell.

Table 1 presents a quantitative comparison of the conversion efficiencies achieved with the proposed optical scheme and those reported for quasi-phase-matched approaches in crystalline quartz.

*Table 1. Comparative analysis of the key parameters of optical schemes for implementing quasi-phase matching in crystalline quartz*

|  | [5] | [6] | Current work |
|---|---|---|---|
| Pump energy, mJ | 1.8 | 2.3 | 3.7 |
| Pulse width, ns | 0.7 | 0.7 | 9.3 |
| Pump spot, um | 120 | 40 | 520 |
| Pump intensity, GW/cm² | 23 | 261 | 0.19 |
| Pump Rayleigh length, mm | 10 | 1 | 200 |
| Second harmonic energy, uJ | 9 | 8 | 1 |
| Conversion efficiency, % | 0.5 | 0.35 | 0.027 |
| Conversion efficiency, %/MW/cm² | $2 \cdot 10^{-5}$ | $1.3 \cdot 10^{-6}$ | $1.4 \cdot 10^{-4}$ |

As shown in Table 1, the absolute conversion efficiency achieved in this work is lower than that reported in the cited studies; however, the efficiency normalized to the pump intensity is significantly higher.

## Conclusion

In this work, we studied the free-space quasi-phase matching in crystalline quartz. We showed that the use of multipass cells can yield an enhancement factor exceeding 1000 as compared to a single pass. This result was obtained at pump intensity of 190 MW/cm², which is three orders of magnitude lower than the damage threshold of crystalline quartz. It opens the way for further scaling up the conversion efficiency. Considering these results, the combination of the described conversion technique with those mentioned in the introduction appears very promising. The proposed scheme is shown in fig.4.

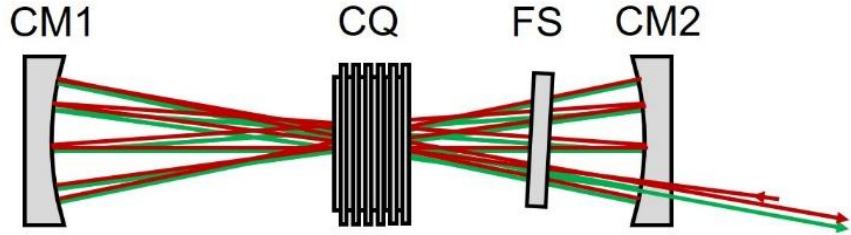

*Fig.4 Proposed optical scheme for combining stacked crystalline quartz and multipass cell: CM1, CM2-curved mirrors, CQ –crystalline quartz stack, FS-fused silica plate for the phase correction.*

We consider the generation of DUV radiation to be a particularly interesting extension of this approach especially in combination with Strontium Tetraborate [10]. The high reflective mirror coatings operating in the DUV range, as well as to anti reflection coatings on the nonlinear crystalline-quartz element are in principle feasible. The only fundamental limiting factor remains the reduction of the coherence length in the UV. We believe that the integration of periodic crystalline-quartz structures into a multipass cell architecture offers considerable potential for strengthening nonlinear optical processes and extending phase-matching capabilities in the deep-ultraviolet spectral range. By enabling higher field intensities and longer effective interaction lengths, this approach could contribute to the development of efficient and scalable sources of coherent DUV radiation.

**Acknowledgment:** The authors would like to thank Kilian Fritsch (n2-Photonics company), Dr. Johann Meyer (Helmut-Schmidt University) and Mitul Akbari for support and helpful discussions.
The work is supported by Deutsche Forschungsgemeinschaft (DFG). Project number 545612524, "Multi-pass cells for next generation laser-plasma accelerators."
**Disclosures:** The authors declare no conflicts of interest.

**Data availability:** Data underlying the results presented in this paper are available upon reasonable request.